# RF-thermal-structural-RF coupled analysis on the travelling wave disk-loaded accelerating structure

PEI Shi-Lun(裴士伦)[1])   CHI Yun-Long(池云龙)   ZHANG Jing-Ru(张敬如)   HOU Mi(侯汨)   LI Xiao-Ping(李小平)

Institute of High Energy Physics, Chinese Academy of Sciences, Beijing 100049, China

**Abstract** Travelling wave (TW) disk-loaded accelerating structure is one of the key components in normal conducting (NC) linear accelerators, and has been studied for many years. In the design process, usually after the dimensions of each cell and the two couplers are finalized, the structure is fabricated and tuned, and then the whole structure characteristics can be measured by the vector network analyzer. Before the structure fabrication, the whole structure characteristics are less simulated limited by the available computer capability. In this paper, we described the method to do the RF-thermal-structural-RF coupled analysis on the TW disk-loaded structure with one single PC. In order to validate our method, we first analyzed and compared our RF simulation results on the 3m long BEPCII structure with the corresponding experimental results, which shows very good consistency. Finally, the RF-thermal-structure-RF coupled analysis results on the 1.35m long NSC KIPT linac accelerating structure are presented.

## 1. Introduction

Accelerating structure is a device used to boost the particle energy. Travelling wave (TW) disk-loaded accelerating structure is mostly used in normal conducting (NC) linear accelerators, such as the BEPC and BEPCII linac [1-2]. In the usual design process, the physical dimensions of each cell and the two couplers are determined first, and then the structure is fabricated and tuned. The whole structure RF characteristics are less simulated but measured after the final microwave tuning, which is because of the structure scale (couples of meters long and centimeters in diameter for S-band) and limited by the available computer capability. However, for the structures with very high RF input power, thermal and structural effects must be studied before the mechanical fabrication, so the travelling wave RF electromagnetic fields in the whole structure need to be calculated first.

Because the characteristics of TW structure with several tens of cells are mainly decided by the regular cells, we propose a method to use redesigned power couplers with azimuth symmetry to replace the original waveguide ones in finite element analysis (FEA). Then the coupled RF-thermal-structural-RF analysis can be done in the multi-physics software package ANSYS [3] with much less computer source used. In order to prove the validity of our methods, we first simulated the RF characteristics of BEPCII linac structure, and then compared with the experimental results, which shows very good consistency except the slight difference of the frequency bandwidth. Finally we investigated the RF-thermal-structure-RF effects in the accelerating structure for NSC KIPT (National Science Center, Kharkov Institute of Physics and Technology, Ukraine) linac [4], which is a cooperative project of ANL (Argonne National Laboratory, USA) and KIPT. The linac is designed by IHEP, and will be used to drive a neutron source on the basis of subcritical assembly [5]. Table 1 shows the main specifications of the TW structures for the BEPCII and KIPT linacs.

Table 1: Main specifications of the TW structures for the BEPCII and KIPT linacs

|  | NSC-KIPT | BEPCII |  |
|---|---|---|---|
| Operation frequency | 2856 | 2856 | MHz |
| Operation temperature | 40.0±0.1 | 45.0±0.1 | °C |
| Number of cells | 34 regular cells 2 coupler cells | 85 regular cells 2 coupler cells |  |
| Section length | 1260 (36 cells) | 3045 (87 cells) | mm |
| Phase advance per cell | $2\pi/3$-mode | $2\pi/3$-mode |  |
| Cell length | 34.989783 | 34.989783 | mm |
| Disk thickness (t) | 5.84 | 5.84 | mm |
| Iris diameter (2a) | 27.887-23.726 | 26.220-19.093 | mm |
| Cell diameter (2b) | 83.968-82.776 | 83.458-81.762 | mm |
| Shunt impedance ($r_0$) | 51.514-57.052 | 53.708-63.294 | MΩ/m |
| Q factor | 13806-13753 | 13783-13711 |  |
| Group velocity ($v_g/c$) | 0.02473-0.01415 | 0.02004-0.0063 |  |
| Filling time | 215 | 823 | ns |
| Attenuation parameter | 0.1406 | 0.5383 | Neper |

## 2. ANSYS simulation methodology

The high frequency, steady state thermal and structural solver modules in ANSYS can be used to do the numerical RF-thermal-structural-RF coupled finite element analysis. By using one program for all the simulations any problems of transferring loads between different softwares were eliminated.

A complete RF-thermal-structural-RF coupled analysis cycle on the TW disk-loaded structure requires 6 steps as outlined below. Here we assume the structure is cooled

1) E-mail: peisl@mail.ihep.ac.cn

with the water cooling jacket, which surrounds the whole accelerating structure axisymmetrically. Currently the ANSYS high frequency module has the limitation that only 3D elements can be used. To minimize the CPU time and memory use, an axis-symmetric 3D model with 1° azimuth angle was created to perform the analysis. Otherwise, the azimuth angle of the model needs to be adjusted correspondingly.

1) With HF119 high frequency tetrahedral element, modal analysis was performed in the vacuum part of each accelerating cell, the cell diameter $2b$ was varied to tune the $2\pi/3$ mode resonating at the nominal operation frequency (2856MHz for the BEPCII and KIPT linacs) by fixing the iris diameter $2a$. Using the obtained dimensions of the first and last cells, the redesigned axisymmetric vacuum models of the input/output couplers can be defined. To minimize the reflection coefficient at the input coupler port to the level of $10^{-2}$ to $10^{-3}$, all of the dimensions shown in Fig. 1 except those related with the middle two cells can be adjusted. Once a flat electric field amplitude distribution along the redesigned coupler's axis is obtained and the phase difference between the middle two cells is ~120° ($2\pi/3$), the coupler's dimension can be finalized.

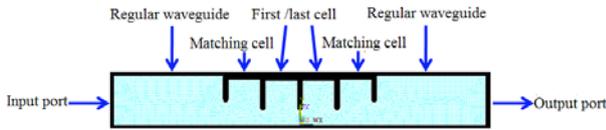

Fig. 1: Schematic of the redesigned input/output coupler

2) Now the vacuum part of the whole structure can be modeled with the obtained cell and coupler dimensions from Step 1. In the meantime, the copper part can also be established with the known dimensions from mechanical design. Both parts need to be meshed with a common surface interface but different 3D elements (HF119 for vacuum part, while SOLID87 for copper part). A common surface mesh can facilitate the load transfer of the RF wall losses to heat flux loads onto the thermal model surfaces.

3) Harmonic analysis in the vacuum part was carried out with the specified average input power. The two redesigned couplers in Step 1 were defined as the power input and output ports, respectively. Impedance boundary condition of copper was applied to the common surface. Using the built-in macro 'SPARM' and 'HFPOWER', it is possible to calculate the scattering (S) parameters and the total time averaged losses. If a harmonic response over a frequency range is performed, the S parameters at each frequency step for each port will be calculated, the structure's bandwidth (VSWR≤1.2) can then be known by checking the S11/VSWR data.

4) With the method described in Ref. [6], the thermal convection coefficient on the cooling water jacket surface can be calculated by the empirical formulae [7]. After the heat flux and the convection coefficient loads were applied on the specific surface of the copper volume, the temperature distribution on the structure wall was calculated.

5) By using the built-in macro 'ETCHG', thermal element SOLID87 was converted to structural element SOLID187. Then the structure distortion caused by RF heating can be simulated by applying appropriate DOF (degree of freedom) constraints and the thermal results as loads.

6) The displacements from the analysis in Step 5 were added to the geometry of the finite element model created in Step 2 with the 'UPGEOM' command. The macro 'ETCHG' was issued again to convert the structural element back to the thermal one. By only selecting the vacuum part model and setting the corresponding RF boundaries, the RF characteristics of the deformed structure can be re-calculated.

It is worth to point out that the mesh size consistency in all of the above analysis steps is the key for acquiring the correct simulation results. Only in Step 1 and 2, the corresponding model dimensions need to be adjusted.

## 3. The BEPCII-linac structure

If we use the real coupler dimensions to simulate the RF characteristics of the whole BEPCII-linac accelerating structure, 1/2 model (180° azimuth angle) needs to be created at least, which is impossible to simulate with one single PC. However, with the method proposed in Section 2, only 1/360 model needs to be established.

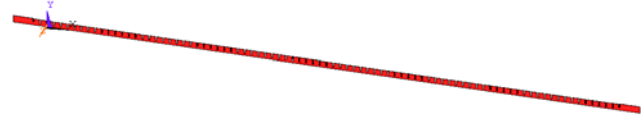

Fig. 2: The 1/360 model of BEPCII linac structure

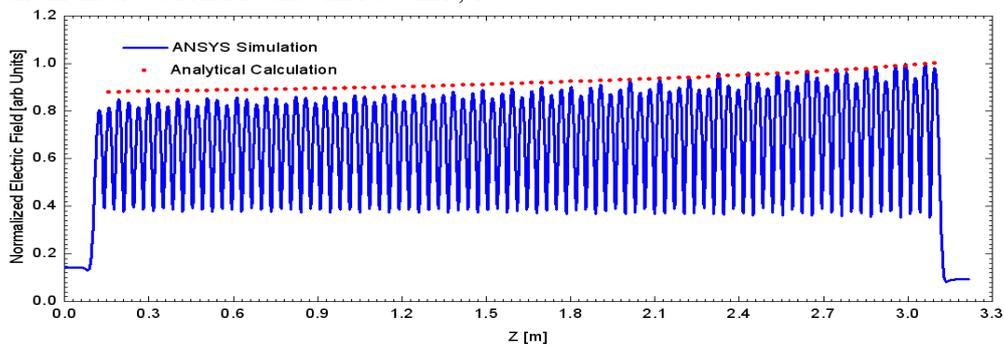

Fig. 3: Amplitude distribution of the electric fields along the structure's axis for the BEPCII-linac

Fig. 2 shows the 1/360 BEPCII linac structure model. Fig. 3 shows the amplitude distribution of the electric field along the structure's axis. The ANSYS result is consistent with the analytical one, which can be calculated with the analytical formulae [8] after obtaining each cell's RF parameters such as the frequency, quality factor $Q_0$, shunt impedance $R$ and the group velocity $v_g$. Usually, the former three parameters can be calculated with SUPERFISH [9], while the last one with MAFIA [10].

With -34dB reflection coefficient at the input port, the ANSYS simulated power attenuation factor to the output port is 0.56Neper (4.85dB), which is very close to the experimentally measured (0.57Neper) and analytically calculated (0.54Neper) ones.

Figs. 4 and 5 show the simulated and experimentally measured VSWR curves respectively. It can be seen that the simulated bandwidth is 2858.05-2854.5=3.55MHz, while the measured one is 2858.42-2853.72=4.7MHz.

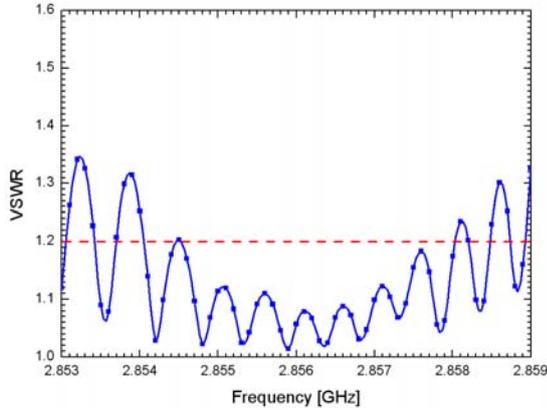

Fig. 4: The ANSYS simulated VSWR curve

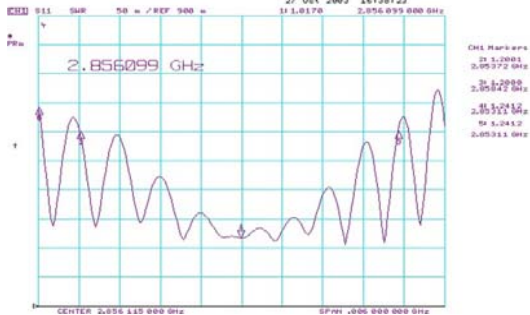

Fig. 5: The measured VSWR curve

Generally speaking, our ANSYS simulation results agree with the experimental ones except the 1.15MHz difference in bandwidth calculation.

## 4. The NSC KIPT-linac structure

Electron with very high beam current (>600mA) will travel through all the 10 accelerating structures of the KIPT-linac [4-5], so very high RF power (20MW for maximum peak power, 20MW×3μs×625Hz=37.5kW for maximum average power) is needed to compensate the beam loading effect [4, 8], water jacket cooling is needed to sufficiently cool down the structure. Due to the energy extraction of beam (beam loading), the electric field in the structure for beam-off case is higher than that for the beam-on case. Since ANSYS cannot simulate the beam loading effect, the RF electric field amplitude and RF heating loss distribution need to be scaled from the beam-off case. The scaling relation can be obtained from the analytical calculation [8].

With the beam-off case, the RF electric field amplitude distribution along the structure's axis is shown in Fig. 6. The attenuation factor simulated by ANSYS is 0.15Neper, while it is 0.14Neper by the analytical calculation.

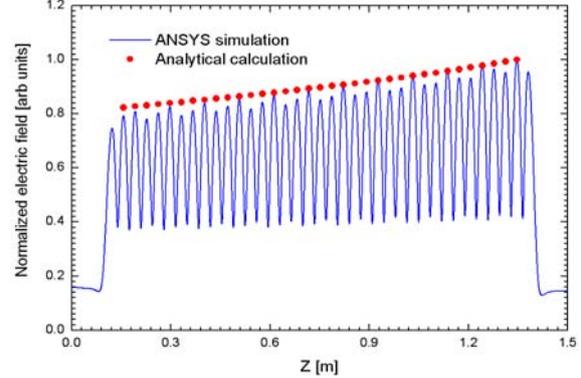

Fig.6: The electric field amplitude distribution along the structure's axis for the KIPT-linac

The cooling water jacket is an annular water pipe with internal and external diameters of 102mm and 116mm. 10t/hour was chosen to be the water flow rate.

When the water cooling temperature is controlled to be 40°C, the temperature distributions along the structure for both the beam-off and beam-on cases are shown in Fig. 7. The maximum temperature rise is located at the output coupler part when the beam is off, which is due to the highest electromagnetic field. On the contrary, when the beam is on, due to the beam loading effect, the input coupler part will have the highest electric field and maximum temperature rise. For both the beam-off and beam-on cases, the highest temperatures are all located at the iris tip parts. The maximum temperature differences are ~10°C and ~7°C for both cases.

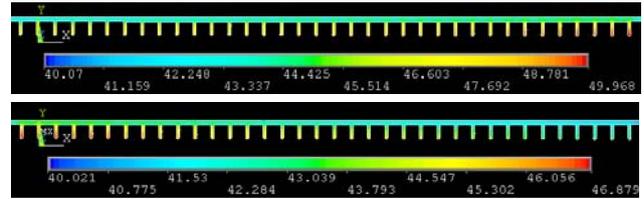

Fig. 7: The temperature distributions along the structure for beam-off (upper) and on (lower) cases (unit: °C)

If the reference temperature is also set to be 40°C, the structure deformations for 40°C cooling water are shown in Figs. 8 and 9. The longitudinal deformation can cause electric field phase shift, while the transversal one is the major source of frequency shift. Here the phase shifts caused by longitudinal deformations are ~0.72° and ~0.60° for the beam-off and beam-on cases, respectively. With the scaling law $\Delta f/\Delta(2b) = -36$kHz/μm [11], the frequency shifts caused by the transversal deformation can be roughly calculated to be ~0.68MHz and ~0.58MHz for both cases.

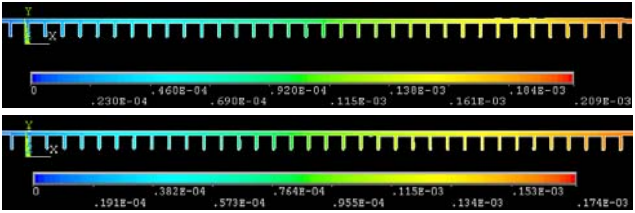

Fig. 8: The longitudinal deformations along the structure for beam-off (upper) and on (lower) cases (Unit: m)

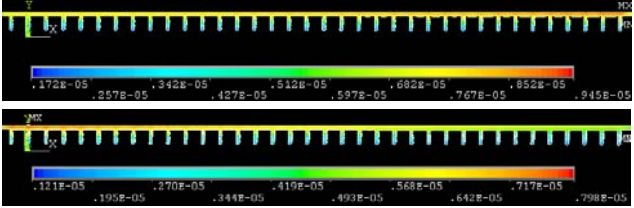

Fig. 9: The transversal deformations along the structure for beam-off (upper) and on (lower) cases (Unit: m)

The VSWRs of the accelerating structure versus frequency for both beam-off and beam-on are shown in Figs. 10 and 11. The RF heating will cause the VSWR curves to shift downward to lower frequencies by ~0.6MHz and ~0.5MHz for both cases, which are a little bit smaller than the scaling law method estimations. This small difference is due to the fact that the scaling law method calculation is estimated by the largest transversal deformation, while in reality not every cell has that large deformation. The bandwidth of the structure is ~5.4MHz defined by VSWR≤1.2, which is fairly larger than the RF heating caused frequency shift and assures the stable operation of the accelerating structure.

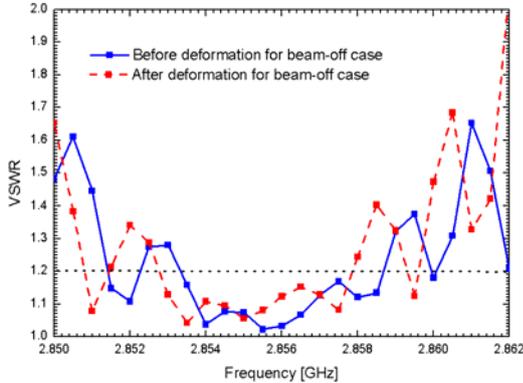

Fig. 10: The VSWR curve for the beam-off case

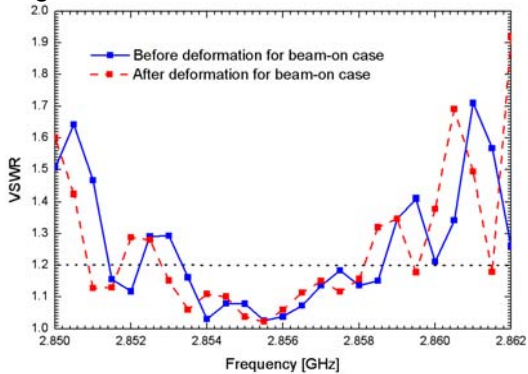

Fig. 11: The VSWR curve for the beam-on case

For the KIPT-linac accelerating structure, once the tuning is finished and the water cooling jacket is mounted, it cannot be tuned any more. At this time, the only way to suppress the RF heating caused structure characteristics deterioration is to adjust and optimize the cooling water temperature.

Figs. 12 to 16 show the structure characteristic variations (such as the maximum temperature, the longitudinal deformation, the total phase shift, the maximum transversal deformation and the frequency shift, etc.) at different cooling water temperatures for both the beam-off and beam-on cases. It can be seen that one can lower the electric field phase shift and the frequency shift by varying the cooling water temperature, and all the listed characteristics in the plots are linearly correlated with the cooling water bulk temperature. The optimized operating water temperature is ~32-33°C.

To check the effect of water flow rate on the structure characteristics' variations, the calculations of 20t/hour water flow rate were also done, the results of which are also shown in Figs. 12 to 16. Compared with the 10t/hour case, its effect on the structure is a little bit smaller, there is about 2°C maximum temperature decrease. However, the cooling water system capability needs to be roughly doubled, so from the cooling power saving point of view, it is not very necessary.

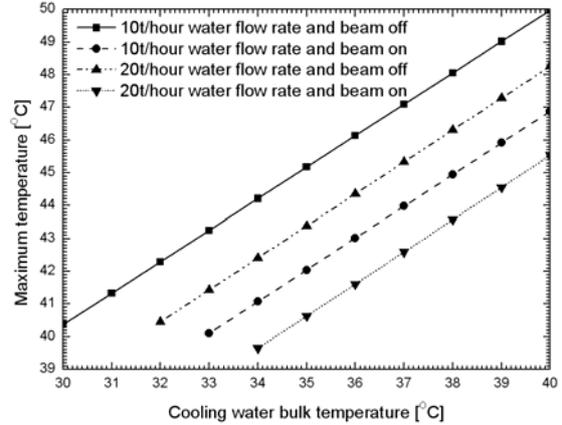

Fig. 12: The maximum temperature on the structure at different cooling water bulk temperatures

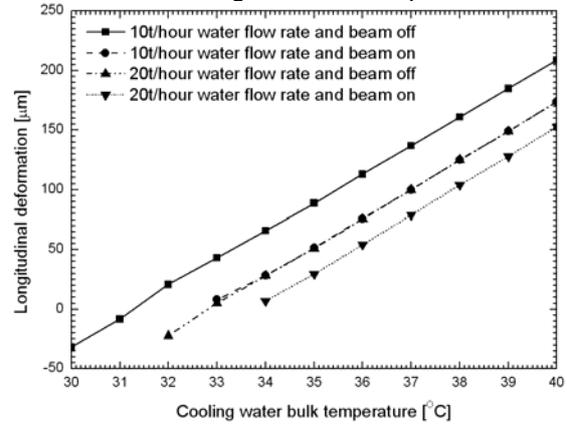

Fig. 13: The longitudinal deformation of the structure at different cooling water bulk temperatures

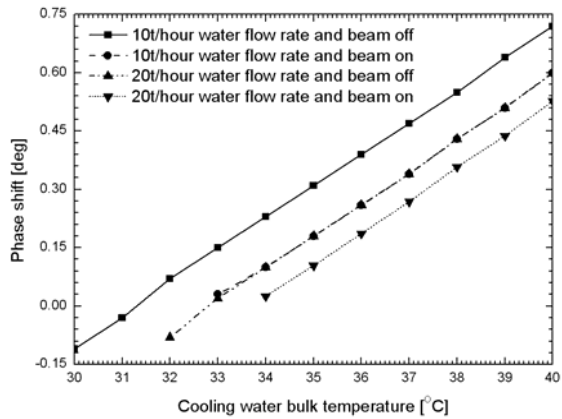

Fig. 14: The total phase shift of the electric field at different cooling water bulk temperatures

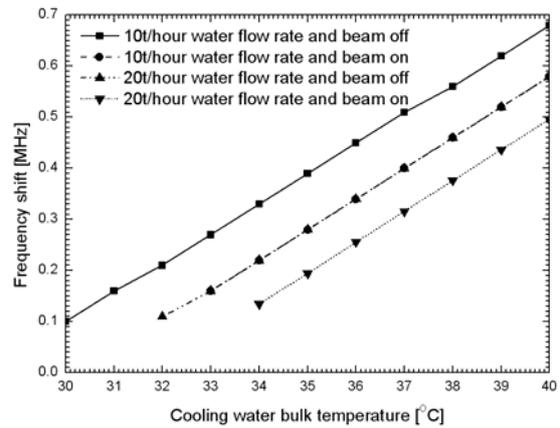

Fig. 16: The frequency shift of the structure at different cooling water bulk temperatures

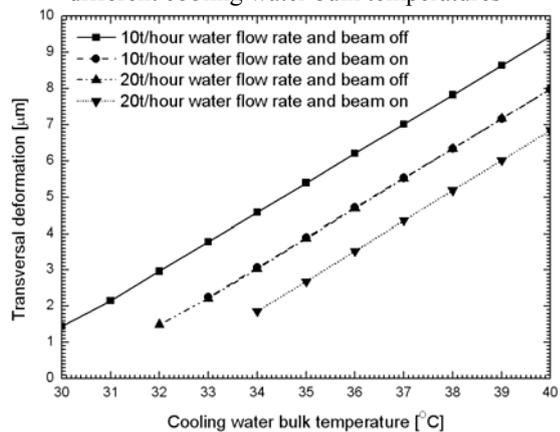

Fig. 15: The transversal deformation of the structure at different cooling water bulk temperatures

## 5. Conclusions

By replacing the real waveguide coupler with the axisymmetric on-axis coupled one in ANSYS, the RF-thermal-structural-RF coupled analysis on the TW disk-loaded accelerating structure can be done with much less computer time and memory used.

The consistency between the simulation and experimental results of the BEPCII linac structure shows the validity of our simulation. The coupled analysis of the KIPT-linac structure shows the optimized cooling water temperature is roughly 32-33$^\circ$C, and the RF heating caused frequency shift is fairly larger than the structure's bandwidth, so 10t/hour water flow rate is enough for stable operation.

## References

[1] The BEPC R & D Report, Part I, Injector, 1989
[2] PEI Guo-Xi et al. Overview of BEPCII Linac Upgrade. Preliminary Design Report of BEPCII-Linac, IHEP-BEPCII-SB-03, 2002.6
[3] ANSYS is a trademark of SAS Inc., (www.ansys.com)
[4] Pei Shi-Lun, Chi Yun-Long et al. Beam Dynamics Studies on the 100MeV/100kW Electron Linear Accelerator for NSC KIPT Neutron Source, Proceedings of IPAC2011, Kursall, San Sebastian, Spain, 2011.9
[5] CHI Yun-Long et al. Design Report on the NSC KIPT Electron Linear Accelerator, Version 2, IHEP, 2011.8
[6] PEI Shi-Lun, WANG Shu-Hong. Chinese Physics C, 2005, 29(9): 912-917
[7] XU Zhao-Jun. Heat Transfer Theory. Beijing: China Machine Press, 1980. 57
[8] YAO Chong-Guo. Electron Linear Accelerator. Beijing: Science Press, 1986
[9] Billen J H, Young L M . Possion Superfish: LA-UR-96-1834. Los Alamos National Laboratory, 2002
[10] The MAFIA Collaboration. MAFIA User Manual (Version 4.106). Germany: CST Inc, 2000
[11] SHU Zhao et al. High Power Laser and Particle Beams, 2011, 23(1): 195-200